\documentclass{elsart}
\usepackage{epsfig}
\usepackage{amssymb}
\journal{Physics Letters B}

\newcommand{\mc}{\multicolumn}
\newcommand{\lsim}{\mathrel{\mathop{\kern 0pt \rlap
  {\raise.2ex\hbox{$<$}}}
  \lower.9ex\hbox{\kern-.190em $\sim$}}}
\newcommand{\gsim}{\mathrel{\mathop{\kern 0pt \rlap
  {\raise.2ex\hbox{$>$}}}
  \lower.9ex\hbox{\kern-.190em $\sim$}}}

\begin{document}
\begin{frontmatter}
\title{Microscopic Approach to Nucleon Spectra in Hypernuclear Non--Mesonic Weak Decay}

\author{E. Bauer$^1$ and G. Garbarino$^2$}

\address{$^1$Departamento de F\'{\i}sica, Universidad Nacional de La Plata and
IFLP, CONICET C. C. 67, 1900 La Plata, Argentina}

\address{$^2$Dipartimento di Fisica Teorica, Universit\`a di Torino, I--10125 Torino, Italy}

\date{\today}

\begin{abstract}
A consistent microscopic diagrammatic approach is applied for the first time
to the calculation of the nucleon emission spectra in the
non--mesonic weak decay of $\Lambda$--hypernuclei.
We adopt a nuclear matter formalism extended to finite nuclei via the local density
approximation, a one--meson exchange weak transition potential
and a Bonn nucleon--nucleon strong potential.
Ground state correlations and final state interactions,
at second order in the nucleon--nucleon interaction,
are introduced on the same footing for all the isospin channels of one-- and two--nucleon
induced decays. Single and double--coincidence nucleon spectra are predicted for
$^{12}_{\Lambda}$C and compared with recent KEK and FINUDA data. The key role played
by quantum interference terms allows us to improve
the predictions obtained with intranuclear cascade codes. Discrepancies with data
remain for proton emission.
\end{abstract}
\end{frontmatter}

Being the only source of information available on
strangeness--changing baryon interactions, the
non--mesonic weak decay of $\Lambda$--hypernuclei
has attracted considerable interest and experienced great advances in the last
years~\cite{Al02,Ga10,Ou05}. These hadronic weak interactions,
whose determination requires the solution of complex many--body problems
also involving strong interaction physics,
are important inputs, for instance, when investigating the
thermal evolution and the stability of compact stars~\cite{Sc10}.
The non--mesonic decay width, $\Gamma_{\rm NM}=\Gamma_1+\Gamma_2$,
is built up from one-- ($1N$) and two--nucleon induced ($2N$) decays,
$\Gamma_{1} = \Gamma_n+\Gamma_p$ and
$\Gamma_{2} = \Gamma_{nn}+\Gamma_{np}+\Gamma_{pp}$, with the
isospin components given by $\Gamma_{N} = \Gamma(\Lambda N \to nN)$
and $\Gamma_{NN'} = \Gamma(\Lambda NN' \to nNN')$, with $N,\, N'=n$ or $p$.

As noted in~\cite{Ba10b}, from a quantum--mechanical point of view
the only observable in non--mesonic weak decay are the
rate $\Gamma_{\rm NM}$ and the spectra of the emitted
nucleons. Each one of the elementary non--mesonic decays occurs in
the nuclear environment, thus final state interactions (FSI) can
modify the quantum numbers of the nucleons produced in such weak
processes and new, secondary nucleons are emitted as well: this prevents
the measurement of any of the non--mesonic partial decay rates.

Although the long--standing puzzles on $\Gamma_n/\Gamma_p$ and the asymmetry parameter
have been solved recently~\cite{Ga10}, a detailed knowledge of the
non--mesonic decay mechanisms is still missing, as demonstrated
by the persistent discrepancies between theory and experiment
on the nucleon spectra~\cite{Ba10}.

In this Letter we present for the first time a consistent microscopic calculation of the
nucleon emission spectra. A nuclear matter formalism is adopted and results
for single and double--coincidence nucleon spectra
are reported for $^{12}_{\Lambda}$C within the local density approximation.
Previous investigations~\cite{Ba07} demonstrated the negligible effect of ring
and RPA microscopic contributions on
nucleon spectra. Ground state correlations
(GSC) and FSI contributions are introduced here at second order in the
nucleon--nucleon interaction for the whole set of
$1N$ and $2N$ isospin decay channels.
The weak transition potential $V^{\Lambda N\to NN}$ contains the exchange of the full set of
mesons of the pseudoscalar ($\pi$, $\eta$, $K$) and vector octets ($\rho$, $\omega$,
$K^*$), with strong coupling constants and cut--off parameters deduced from
the Nijmegen soft--core interaction NSC97f~\cite{St99};
for the nucleon--nucleon interaction $V^{NN\to NN}$ we adopt the Bonn potential
(with the exchange of $\pi$, $\rho$, $\sigma$ and $\omega$ mesons) \cite{Ma87}.
Being fully quantum--mechanical, the present approach is expected to produce
more reliable results for FSI than those
based on the (semi--classical) nucleon rescattering given by
intranuclear cascade (INC) models~\cite{Ra97,Ga03}.

\begin{figure}[t]
\begin{center}
\mbox{\epsfig{file=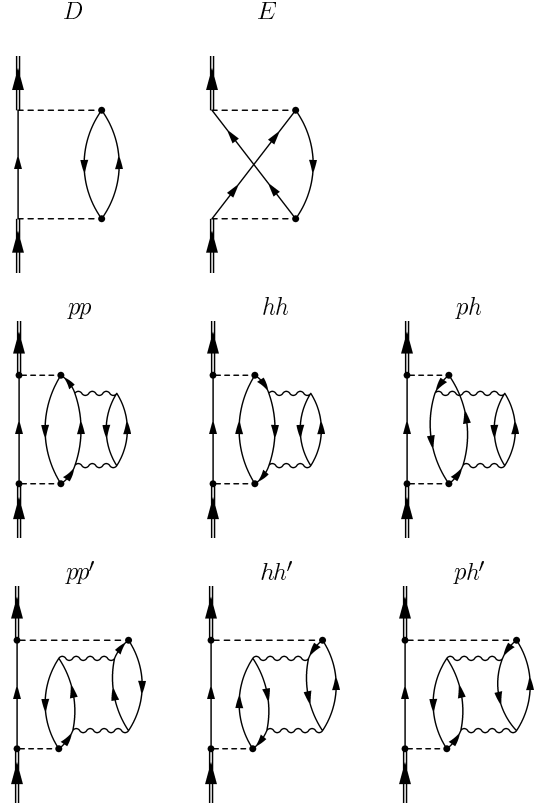,width=.5\textwidth}}
\caption{The set of Feynman diagrams considered in this work for the in--medium $\Lambda$
self--energy. $D$ and $E$ are the direct and exchange terms of the $1N$ decay
channel, the remaining ones contribute to $2N$ and FSI--induced decays.
The dashed and wavy lines stand for the weak and strong
potentials, $V^{\Lambda N \to NN}$ and $V^{NN\to NN}$, respectively.}
\label{fig1}
\end{center}
\end{figure}

\begin{figure}[h]
\begin{center}
\mbox{\epsfig{file=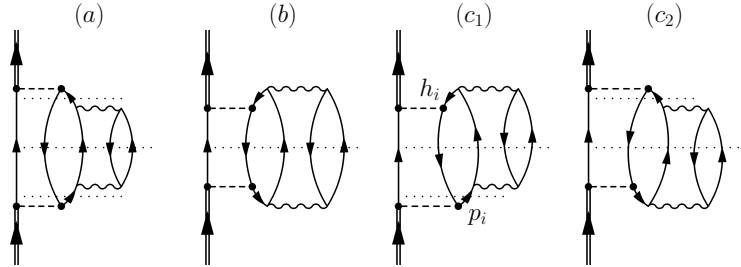,width=.7\textwidth}}
\caption{The Goldstone diagrams corresponding to the
$pp$ Feynman diagram of Fig.~\protect\ref{fig1}.
Note that the QIT ($c_1$) and ($c_2$) are vanishing since $p_i=h_i$
and then one of the two requirements on the Fermi level,
$p_i>k_F$ and $h_i\leq k_F$, cannot be fulfilled.
See text for details.}
\label{fig2}
\end{center}
\end{figure}

The many--body terms we consider are represented by the in--medium
$\Lambda$ self--energy Feynman diagrams of Fig.~\ref{fig1}, where
we limit to two-- and three--nucleon emission.
Diagrams $D$ and $E$ contribute to $1N$ decays.
By considering all the possible time--orderings of the
Feynman diagrams at second order in $V^{NN\to NN}$, one obtains Goldstone diagram
contributions to $2N$ and FSI--induced decays as well as quantum interference terms (QIT).
FSI Goldstone diagrams have at least one $V^{N N\to NN}$
acting after $V^{\Lambda N\to NN}$.
QIT always involve a FSI and are of two kinds: FSI amplitudes interfere
with both $1N$ ($1N$--FSI QIT) and $2N$ decay amplitudes ($2N$--FSI QIT).
We shall use the term 'plain FSI' to indicate a FSI which is
not a QIT.
For example, from the $pp$ Feynman diagram of Fig.~\ref{fig1}
one obtains the Goldstone diagrams of Fig.~\ref{fig2}:
diagram (a) provides a plain FSI contribution when a cut on $3p2h$ states is considered
(this cut is associated to three--nucleon emission), while
for cuts on $2p1h$ states (two--nucleon emission) one has $1N$--FSI QIT; (b) is a
$2N$ decay contribution, introduced by GSC; ($c_1$) and ($c_2$) are
QIT with two possible final states, $2p1h$,
contributing to $1N$--FSI QIT, and $3p2h$, leaving to $2N$--FSI QIT
(incidentally, they are vanishing since
$p_i=h_i$ and then one of the two requirements on the Fermi level,
$p_i>k_F$ and $h_i\leq k_F$, cannot be fulfilled).

From this analysis it is clear that a single Goldstone
diagram can contribute to a plain FSI and to a QIT, depending on the
final physical state one considers:
this point is important to have an unambiguous interpretation of the QIT.
A cut in a Goldstone diagram that leads to two symmetric
pieces is interpreted as the square of a transition amplitude:
a non--QIT which contains the interaction $V^{NN\to NN}$ can represent
either a $2N$ decay or a plain FSI--induced decay.
At variance, when the cut divides the diagram into
two different amplitudes, one has a QIT. $2N$ and plain FSI contributions
are positive--definite, while QIT can be either positive or negative.
This trivial fact has important consequences that will be discussed later.

Regardless of the set of many--body contributions one considers,
the Goldstone diagram technique allows us to write the total number of
nucleons and nucleon pairs emitted in the non--mesonic decay
as follows~\cite{Ba07}:
\begin{eqnarray}
\label{nn1f}
N_{n} &= & 2 \bar{\Gamma}_{n} +  \bar{\Gamma}_{p} +
3 \bar{\Gamma}_{nn} + 2 \bar{\Gamma}_{np} + \bar{\Gamma}_{pp}
+ \sum_{i, \, f} N_{f \, (n)}
\bar{\Gamma}_{i, f}~, \nonumber \\
\label{np1f}
N_{p} & = & \bar{\Gamma}_{p} + \bar{\Gamma}_{np} + 2
\bar{\Gamma}_{pp} +
\sum_{i, \, f} N_{f \, (p)} \, \bar{\Gamma}_{i, f}~,\nonumber \\
\label{nnn1f}
N_{nn} & = & \bar{\Gamma}_{n} + 3 \bar{\Gamma}_{nn}
+ \bar{\Gamma}_{np}+
\sum_{i, \, f} N_{f \, (nn)} \, \bar{\Gamma}_{i, f}~,\nonumber \\
\label{nnp1f}
N_{np} & = & \bar{\Gamma}_{p} + 2 \bar{\Gamma}_{np}
+ 2 \bar{\Gamma}_{pp} +
\sum_{i, \, f} N_{f \, (np)} \, \bar{\Gamma}_{i, f}~,\nonumber \\
\label{npp1f}
N_{pp} & = & \bar{\Gamma}_{pp} +
\sum_{i, \, f} N_{f \, (pp)} \, \bar{\Gamma}_{i, f}~,
\end{eqnarray}
where a normalization per non--mesonic decay is used
($\bar{\Gamma} \equiv \Gamma/\Gamma_{\rm NM}$).
Single and double coincidence nucleon spectra are obtained by constraining the
evaluation of each $\bar \Gamma$ to certain intervals in energy, opening angle, etc.
The $\bar \Gamma_N$'s ($\bar \Gamma_{NN'}$'s)
are the $1N$ ($2N$) decay rates, while
the remaining terms containing the functions $\bar{\Gamma}_{i, f}$
represent FSI Goldstone diagrams
(for instance, diagrams (a), ($c_1$) and ($c_2$) of Fig.~\ref{fig2}).
The index $i$ in $\bar{\Gamma}_{i, f}$ is used to label the FSI Goldstone
diagrams obtained from the
Feynman diagrams of Fig.~\ref{fig1} at second order in $V^{NN\to NN}$,
while $f$ denotes the final physical states of the Goldstone diagram and in the
present case can take the values $f=nN$ (cut on $2p1h$ states) and $nNN'$ (cut on $3p2h$
states). Finally, $N_{f \, (N)}$ ($N_{f \, (NN')}$) is the number of
nucleons of the type $N$ (of $NN'$ pairs) contained in the multinucleon state $f$.

Note that Eqs.~(\ref{npp1f}) contain five observables, $N_{N}$ and $N_{NN'}$, and
five unobservable decay widths, $\Gamma_{N}$ and $\Gamma_{NN'}$.
The unobservable character of the decay widths is due to
the presence of FSI, which we know to be important.
Therefore, one cannot invert the above relations to obtain the
decay widths in terms of the observed spectra. The experimental values of the
partial widths can be obtained from the measured spectra
only if a model is used for deconvoluting the FSI effects.

To illustrate an important property of FSI,
let us consider the diagram of Fig.~\ref{fig2}(a) (we assign to it the
index $i=1$) with $f=nNN'$ (cut on $3p2h$ states): it is
represented by the function $\bar{\Gamma}_{1, nNN'}$, which is a plain
FSI--induced contribution.
The diagram \ref{fig2}(a) can be cut also on $2p1h$
states: in this case it contributes to the $1N$--FSI QIT $\bar{\Gamma}_{1, nN}$.
From sum rule considerations on Goldstone diagrams \cite{Alberico} it follows that
\begin{equation}
\sum_{N=n,p}\bar\Gamma_{1, nN} + \sum_{N,N'=n,p}\bar\Gamma_{1, nNN'} = 0.
\label{sr1}
\end{equation}
In general, each FSI $\Lambda$ self--energy diagram
can in no way contribute to the total non--mesonic rate:
\begin{equation}
\sum_{N=n,p}\bar\Gamma_{i, nN} + \sum_{N,N'=n,p}\bar\Gamma_{i, nNN'}
+\sum_{N,N',N''=n,p}\bar\Gamma_{i, nNN'N''}+\dots= 0,
\label{sr2}
\end{equation}
for any value of $i$, while in principle it
can affect the $1N$, $2N$, etc, decay rates separately
(see however the discussion of~\cite{Ba10b} on the opportunity not to include
these FSI terms in the definition of the unobservable partial rates).
Note that the above sum rule is valid for the nucleon
spectra integrated over the whole energy range for all the nucleons in the
state $f$; the cancelation between the $3p2h$ and $2p1h$ contributions of
each Goldstone diagram does not occur at a given value of the energy of one
of the nucleons contained in $f$.
Note also that the numbers $N_{f \, (N)}$ and $N_{f \, (NN')}$ prevent the above
cancelation to occur for the total numbers in Eqs.~(\ref{npp1f}) too.
Thus, QIT diagrams surely play a role in nucleon spectra.

Let us emphasize some consequences of the sum rule condition in Eq.~(\ref{sr2}).
First, while $\Gamma_{n}$ and $\Gamma_{p}$ are not observable,
the total decay width $\Gamma_{\rm NM}$ (which is unaffected by FSI) is an observable.
Second, it is obvious that the cancellation in
Eqs.~(\ref{sr1}) and (\ref{sr2}) occurs because some terms are
negative: these terms are necessarily QIT. It is well known
that FSI are important; in addition the second property above tells us
that the magnitude of
QIT is the same as the one of plain FSI in the evaluation of $\Gamma_{\rm NM}$.
This proportion can change in the evaluation of the spectra due
to the presence of the numbers $N_{f \, (N)}$ and $N_{f \, (NN')}$,
but in no way QIT can be neglected: without QIT,
the emission spectra would be overestimated.
It is important to stress that this discussion refers to a
quantum mechanical treatment only.


Before discussing our numerical results, it should be mentioned that
the values of $\bar \Gamma_N$, $\bar \Gamma_{NN'}$ and $\bar{\Gamma}_{i, f}$
are obtained from the corresponding $\Lambda$ self--energy Goldstone diagrams.
Explicit expressions for diagrams $D$ and $E$ in
Fig.~\ref{fig1} are found in~\cite{Ba03}.
For the formal derivation of the decay rates from the
$pp$, $ph$, $hh$, $pp'$, $ph'$ and $hh'$
Feynman diagrams in the same figure,
we proceed as follows. The evaluation of the
$pp$, $ph$ and $hh$ Goldstone diagrams contributing to $2N$ decays
can be found in~\cite{Ba04} ($pp$ in the
main text, the other two in the Appendix). The expressions for the
same time--orderings for $pp'$, $ph'$ and $hh'$ diagrams
are shown in the Appendix of~\cite{Ba09b}.
The remaining time--ordering formulas for $pp$, $ph$, $hh$, $pp'$, $ph'$ and $hh'$
can be obtained from the just mentioned ones after some work:
in any case, the main ingredients are already present in the
mentioned works.

As a starting point for the numerical analysis, in
Table~\ref{gammas} we show the decay rates obtained for $^{12}_\Lambda$C.
We also predict that $\Gamma_{np}/\Gamma_{2}=0.84$,
$\Gamma_{pp}/\Gamma_{2}=0.12$ and $\Gamma_{nn}/\Gamma_{2}=0.04$.
The agreement with the recent KEK and FINUDA data is quite satisfactory.
\vspace{0.3cm}
\begin{table}[h]
\begin{center}
\caption{The non--mesonic decay widths predicted for $^{12}_\Lambda$C
(in units of the free decay rate). The most recent data, from KEK--E508~\protect\cite{Kim09}
and FINUDA~\protect\cite{FINUDA}, are also given.}
\label{gammas}
\begin{tabular}{l c c c}\hline\hline
\mc {1}{c}{Decay rate} &
\mc {1}{c}{Our} &
\mc {1}{c}{KEK--E508} &
\mc {1}{c}{FINUDA}\\ \hline
$\Gamma_n$                 & $0.15$ & $0.23\pm 0.08$ &\\
$\Gamma_p$                 & $0.47$ & $0.45\pm 0.10$ &\\
$\Gamma_1$                 & $0.62$ & $0.68\pm 0.13$ &\\
$\Gamma_2$                 & $0.36$ & $0.27\pm 0.13$ &\\
$\Gamma_{\rm NM}$          & $0.99$ & $0.95\pm 0.04$ &\\
$\Gamma_n/\Gamma_p$        & $0.33$ & $0.51\pm 0.13\pm 0.05$ &\\
$\Gamma_2/\Gamma_{\rm NM}$ & $0.37$ & $0.29\pm 0.13$  & $0.24\pm 0.10$\\
\hline\hline
\end{tabular}
\end{center}
\end{table}

We turn now to our main concern in this Letter, which is the study of the
emission spectra. The obtained emission spectra depend mainly on the phase
space and the nucleon--nucleon strong potential. The fact that these spectra
are normalized per non--mesonic weak decay makes the dependence on the weak
interaction to be very small. In the present
contribution we have employed the weak and strong interaction models used
in~\cite{Ba10b}. These
parameterizations turned out to reproduce the full set of decay widths for
carbon hypernuclei determined recently at KEK~\cite{Kim09}. Note that the
use of the same strong interaction model for both GSC and FSI
not only provides internal consistency to the model, but gives us a
criterion to select the strong interaction potential, since modifications of
this potential leads to testable changes in both $\Gamma_2/\Gamma_{\rm
NM}$ and the nucleon spectra.

\begin{figure}[h]
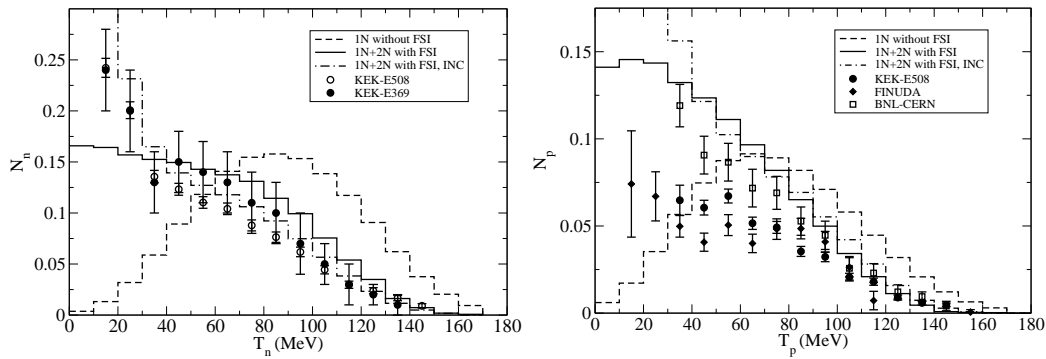

\begin{center}
\mbox{\epsfig{file=NnTn.eps,width=.49\textwidth}}
\mbox{\epsfig{file=NpTp.eps,width=.49\textwidth}}
\caption{Neutron and proton kinetic energy spectra for $^{12}_\Lambda$C non--mesonic
weak decay. The dashed (continuous, dot--dashed) lines are normalized
per $1N$ decay (per non--mesonic decay). Experimental data are from
KEK--E369~\protect\cite{Kim03}, KEK--E508~\protect\cite{Ok04},
FINUDA~\protect\cite{FINUDA} and BNL--CERN~\protect\cite{Mo74}.}
\label{NpTp}
\end{center}
\end{figure}

In Fig.~\ref{NpTp} we show the neutron and proton
kinetic energy spectra for the non--mesonic decay
of $^{12}_\Lambda$C. The dashed curves are the distributions
of the $1N$ decay nucleons (normalized per $1N$ decay):
as expected, they show a maximum at half of the $Q$--value
for $^{12}_\Lambda$C non--mesonic decay and a bell--type shape
due to the nucleon Fermi motion and the $\Lambda$ momentum distribution
in the hypernucleus. The inclusion of $2N$ and FSI--induced decay processes
provides the results given by the continuous lines
(normalized per non--mesonic decay)
and leads to a reduction of the nucleon average energy,
thus filling the low--energy part of the spectrum and emptying the high--energy
region. This outcome is explained as follows.
The dominant contribution to FSI is provided by diagram~\ref{fig2}(a):
according to phase space arguments,
when a cut on $3p2h$ states is considered one obtains a distribution that
decreases monotonically with $T_N$, while a cut on $2p1h$ states provides a bell--shaped,
negative QIT peaked at $T_N\simeq 70$ MeV. Another important contribution
is that of diagram ~\ref{fig2}(b), which admits $3p2h$ final states only: it
produces positive and monotonically decreasing nucleon distributions with
the same order of magnitude of the $3p2h$ term of diagram ~\ref{fig2}(a).
While we reproduce fairly well the KEK neutron spectra,
a rather strong overestimation is found of the
KEK--E369 and FINUDA proton distributions. Our proton spectrum is instead closer to the
old BNL--KEK data.

Also from Fig.~\ref{NpTp} we see a certain dispersion among the
various experimental proton spectra (for instance, FINUDA shows
a peaking structure at $T_p\simeq 80$ MeV which is not seen in the other
data nor in this and in previous calculations). Compared with the nucleon spectra
obtained with the INC model for FSI in~\cite{Ba10} (dot--dashed curves),
we obtain here similar
results for $T_N \gsim 40$ MeV; the distributions of~\cite{Ba10} for smaller
$T_N$ increase well beyond the predictions of the present work,
which are more reliable than the ones based on the INC in this energy regime.
Given the obvious differences between the present microscopic approach and the
INC rescattering model, the previous agreement is a remarkable result.
The QIT of the present approach replace the nucleon rescattering present in the
INC, whose effect is to reduce the energy of the nucleons propagating through the
residual nucleus.

\begin{figure}[h]
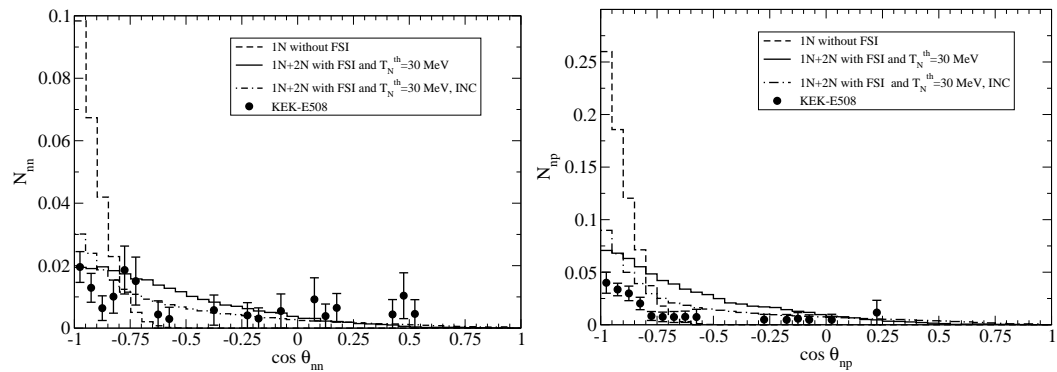

\begin{center}
\mbox{\epsfig{file=Nnnc.eps,width=.49\textwidth}}
\mbox{\epsfig{file=Nnpc.eps,width=.497\textwidth}}
\caption{Opening angle distribution of $nn$ and $np$ pairs.
Normalization is as in Fig.~\protect\ref{NpTp}. Data are from
KEK--E508~\protect\cite{Kim06}.}
\label{Nnpc}
\end{center}
\end{figure}

The opening angle distributions of $nn$ and $np$ pairs
are reported in Fig.~\ref{Nnpc}.
To adhere to the KEK data, the predictions of the full model are
obtained for a 30 MeV nucleon kinetic energy threshold $T^{\rm th}_N$.
The distributions from the $1N$ decay (dashed curves)
are strongly peaked at $\theta_{NN'}=180^\circ$.
The QIT again have a crucial effect: they considerably reduce the back--to--back
contribution, thanks mainly to diagram~\ref{fig2}(a) with $2p1h$ final states;
the non back--to--back region is strongly populated
mainly by the $3p2h$ contributions of diagrams~\ref{fig2}(a)
and \ref{fig2}(b) (see continuous curves).
The final results for the angular correlations turn out to be somewhat less
back--to--back peaked than what found in~\cite{Ba10} (dot--dashed curves),
and, as in that work, are very sensitive to the value adopted for $T^{\rm th}_N$.
The agreement with KEK--E508 data is rather good for the $nn$ spectrum,
while for $np$ pairs a significant overestimation is obtained. Note that we
reproduce for the first time the experimental $nn$ opening angle spectrum: it is the
quantum--mechanical nature of our scheme, i.e., the relevance of QIT, which
brings to this achievement.
Our overestimation of the KEK--E508 $np$ distribution is compatible
with the overestimation of the proton spectrum obtained in the same experiment
(see Fig.~\ref{NpTp}).

\begin{figure}[h]
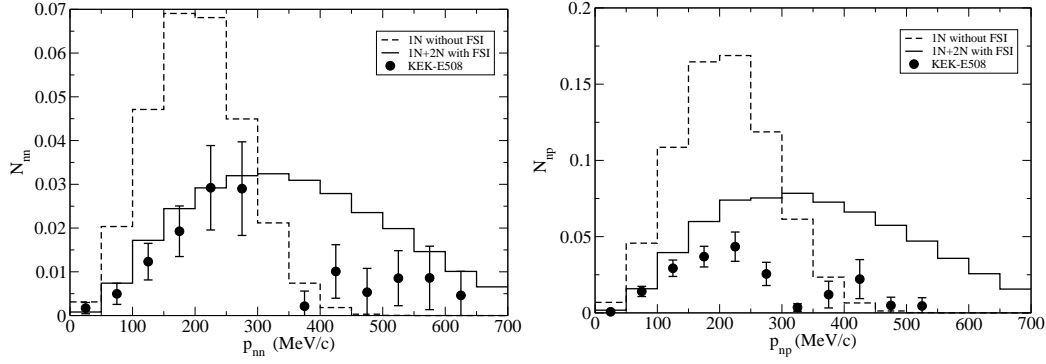

\begin{center}
\mbox{\epsfig{file=Nnnp12.eps,width=.49\textwidth}}
\mbox{\epsfig{file=Nnpp12.eps,width=.49\textwidth}}
\caption{Momentum correlation spectra of $nn$ and $np$ pairs,
with $p_{NN'}\equiv |\vec{p}_{N}+\vec{p}_{N'}|$.
Normalization is as in Fig.~\protect\ref{NpTp}. Data are
from KEK--E508~\protect\cite{Kim09}.}
\label{Nnpp12}
\end{center}
\end{figure}

In Fig.~\ref{Nnpp12} we give the two--nucleon momentum correlation spectra,
i.e., the ${nn}$ and ${np}$ distributions as a function of the
momentum sum $p_{NN'}\equiv|\vec{p}_{N}+\vec{p}_{N'}|$
of two of the outgoing nucleons.
The dashed lines correspond to the $1N$ decay,
while the continuous curves show the full result, with $1N$, $2N$
and FSI included together with a nucleon kinetic energy threshold
$T^{\rm th}_N=30$ MeV, as in the data also shown in the figures.
As noted in Ref.~\cite{Kim09}, the minimum in both the ${nn}$ and ${np}$
KEK--E508 distributions is mainly due to the low statistics
and detection efficiency for events with $p_{NN'}\gsim$ 350 MeV/c
(the KEK detector geometry being optimized for back--to--back coincidence events);
indeed, such dip structure has not been found in our calculation,
which overestimates the data for large correlation momenta
(especially for $N_{np}$, consistently with the spectra discussed so far).

\begin{figure}[h]
\begin{center}
\mbox{\epsfig{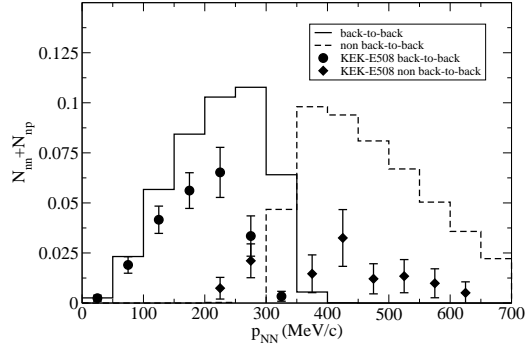}}
\caption{Momentum correlation spectra for the sum of the
$nn$ and $np$ pair numbers for the back--to--back
($\cos \theta_{NN}<-0.7$, continuous line) and non back--to--back
kinematics ($\cos \theta_{NN}>-0.7$, dashed line).
Normalization is per non--mesonic weak decay.
Data are from KEK--E508~\protect\cite{Kim10}.}
\label{Nnnnpp12}
\end{center}
\end{figure}

The distributions of Fig.~\ref{Nnpp12} at low momentum sum
(say below 400 MeV/c) are mainly due to $1N$ decays
(which are strongly back--to--back correlated), while for higher momenta
the contribution of  $2N$ and FSI--induced decays
is dominant (and produces less back--to--back correlated pairs).
This behavior is confirmed by the momentum correlation of the sum
$N_{nn}+N_{np}$ shown in Fig.~\ref{Nnnnpp12}
for the opening angle regions with $\cos \theta_{NN'}<-0.7$
and $\cos \theta_{NN'}>-0.7$.

In Table~\ref{number-ratio} we show results for the coincidence numbers
$N_{nn}$, $N_{np}$ and $N_{pp}$ obtained for  $T^{\rm th}_N= 30$ MeV
and two angular regions, $\cos \theta_{NN'}<-0.7$ and $\cos \theta_{NN'}>-0.7$.
Previous predictions of finite nucleus~\cite{Ga03} and nuclear matter
approaches~\cite{Ba10}, both based on the INC,
and experimental data are also given. Again, a remarkable result is the fact that
the present predictions are very similar to the ones of~\cite{Ba10}, while
one notes less agreement with~\cite{Ga03}, mainly due to
the different models adopted to describe $2N$ decays. Comparison with data
shows an overestimation of $N_{np}$ and $N_{pp}$ but rather good
results for $N_{nn}$. This confirms a systematic overestimation of the
proton emission reported by KEK--E508.
\vspace{0.3cm}
\begin{table}[h]
\begin{center}
\caption{The nucleon coincidence numbers
are given for $T^{\rm th}_N=30$ MeV and
the angular regions with $\cos\, \theta_{NN'}< -0.7$
and $\cos\, \theta_{NN'}> -0.7$ (in parenthesis).
Data are from KEK--E508~\protect\cite{Kim06}.}
\label{number-ratio}
\begin{tabular}{l c c c} \hline\hline
\mc {1}{c}{} &
\mc {1}{c}{$N_{nn}$} &
\mc {1}{c}{$N_{np}$} &
\mc {1}{c}{$N_{pp}$} \\ \hline
This work        & $0.11\,\, (0.13)$ & $0.35\,\, (0.31)$ & $0.04\,\, (0.05)$ \\
INC \cite{Ga03}  & $0.15\,\, (0.18)$ & $0.35\,\, (0.52)$ & $0.08\,\, (0.27)$ \\
INC \cite{Ba10}  & $0.11\,\, (0.10)$ & $0.30\,\, (0.25)$ & $0.05\,\, (0.07)$ \\ \hline
KEK--E508        & $0.083\pm 0.014$ & $0.138\pm 0.014$ & $0.005\pm 0.002$ \\
                 & $(0.083\pm 0.020)$ & $(0.060\pm 0.018)$ & \\
\hline\hline
\end{tabular}
\end{center}
\end{table}

Summarizing, a microscopic approach including GSC and FSI on the same footing
is used to evaluate the nucleon emission spectra in non--mesonic weak decay of
hypernuclei. Within our microscopic model, QIT play a key role: in the
single--nucleon emission spectra they are responsible for moving intensity from
the high--energy region to the low--energy region, while in the opening angle
distribution of nucleon pairs the strong reduction of the back--to--back
peak is entirely due to QIT.
Discrepancies with experiment remain, but are relegated to spectra involving protons.
Further work is in order to understand such a disagreement.
A forthcoming coincidence experiment
at J--PARC~\cite{jparc} will allow a measurement of the nucleon spectra
and a determination of $\Gamma_n$, $\Gamma_p$ and $\Gamma_2$ for $^{12}_\Lambda$C
with improved accuracy. Moreover, our
microscopic approach is particularly suitable for the inclusion of new decay mechanisms
introduced by fermion antisymmetrization and the $\Delta(1232)$--resonance.
These contributions deserve consideration.

\end{document}